# An assessment of European electricity arbitrage using storage systems

Fernando Núñez, David Canca, Ángel Arcos-Vargas [1]

School of Engineering
Department of Industrial Engineering and Management Science.
University of Seville. Spain.

**Abstract.**
Interest in the arbitrage business using storage systems has increased in recent years due to the fast-technological changes in both power electronics and electrical storage. This study analyses the current viability of this business based on a sample of European countries in the year 2019; countries where electricity prices (day-ahead market) and financial conditions show a certain degree of heterogeneity. We basically follow a sequence of three analyses in our study. Firstly, a Linear Mixed-Integrated Programming model has been developed to optimize the arbitrage strategy for each country in the sample. We have considered a 10 MWh Li-Ion battery with 92% round trip efficiency and an expected life of 5000 cycles. This battery can also be combined with different converter sizes, from 1 to 10 MW. Secondly, using the cash-flows from the optimization model, we calculate two financial indicators (NPV and IRR) in order to select the optimal converter size for each country. Tax and discount rates specific to each country have been used with the calculation of this second rate following the methodology proposed by the Spanish regulator. Thirdly, a mixed linear regression model is proposed in order to investigate the importance of observed and unobserved heterogeneity (at country level) in explaining the business profitability. The findings show that, in the near future, the most attractive European countries for the electricity arbitrage business should be the United Kingdom and Ireland, with current NPV close to –400,000 €, while Spain and Portugal might show the worst performances, their current NPV are close to –800,000 €. It can also be observed that, once we control for the investment costs and for the existence of unobservable heterogeneity at country level, the variables which most affect the business profitability are the fiscal requirements in each country and the difference between the maximum and minimum daily electricity prices, especially the standard deviation of this price gap.

**Highlights.**
- European countries show significant differences in wholesale electricity prices and financial conditions.
- A Mixed-Integer Linear Programming model allows the generation of optimal arbitrage.
- The determinants of the financial indicators in each country are analysed with a multilevel model.
- The business is getting closer to being profitable in some European countries.

---

[1] Corresponding author: aarcos@us.es

**Nomenclature**

*Abbreviations*

| | |
|---|---|
| BESS | Battery Energy Storage System |
| CAES | Compressed Air Energy Storage |
| CES | Community Energy Storage |
| DSO | Distribution System Operator |
| EA | Energy Arbitrage |
| EA-PS | Energy Arbitrage - Peak Shaving |
| EDLC | Ultra capacitors |
| EES | Electric Storage System |
| GenCo | Generation Company |
| IRR | Internal Tase of Return |
| LA | Lead-acid |
| Li-Ion | Lithium - Ion |
| LV | Low Voltage |
| NMC | Lithium Nickel Manganes Cobalt |
| NPV | Net Present Value |
| NYISO | The New York Independent System Operator |
| OMIE | Iberian market operator |
| PHS | Pumping Hydraulic System |
| Sna | Sodium-Sulfur |
| TSO | Transmission System Operator |
| VRFB | Vanadium Redox Flow Battery |
| ZEBRA | Nickel-Sodium Chloride Battery |

*Parameters*

$T$ Set of periods (Hours) defining the planning horizon.

$\omega_t$ Market electricity price at hour $t \in T$.

$\psi_o$ A parameter used to define the initial battery capacity (we will consider a battery of 10 MWh in our experiments).

$\varphi$ Deterioration of the battery capacity per cycle. We consider that after 5000 cycles the initial battery capacity decreases by up to 20% of the initial one.

$\psi_{max}$ Conversion rate of the converter.

$\lambda$ A parameter that defines the loss of energy in charging and discharging conversion processes. This parameter is measured as a percentage of the energy purchased or sold.

*Variables*

$\alpha_t$ Binary variable. It takes value 1 if the battery is charged at period (hour) $t \in T$, 0 otherwise.

$\beta_t$ Binary variable. It takes value 1 if the battery is discharged at period $t \in T$, 0 otherwise.

$\delta_t$ Binary variable. It takes value 1 if the battery has finished a charge\discharge cycle at period $t \in T$, 0 otherwise. They are in charge of measuring the battery deterioration with use.

$\gamma_t$ Binary variable. It takes value 1 if the last performed operation prior to period $t \in T$ was a charge and 0 if it was a discharge. These variables will be used as auxiliary variables to enforce cycle occurrences.

$P_t$ Represents the amount of electricity purchased and stored in the battery during period $t \in T$. The purchase decision at period $t$ is conditioned by $\alpha_t$.

$S_t$ Measures the amount of energy discharged from the battery during period $t \in T$. This energy is transformed into electricity and sold at the same period. The sale decision at period $t$ is conditioned by $\beta_t$.

$C_t$ Integer variable that measures the cumulative number of charging\discharging cycles that has been done just until period $t \in T$.

$K_t$ Represents the remaining capacity of the battery at the end of period $t \in T$.

$L$ Measures the level of energy in the battery at the end of period $t \in T$.

# 1. Introduction.

The present study attempts to show that, in addition to the necessary investment and the technical characteristics of the electrical storage, the effectiveness of electricity arbitrage in Europe is strongly influenced both by daily price variations –differences between maximum and minimum prices throughout the day– and by the level of financial risk faced by each national electricity market. We follow the research line of Arcos-Vargas *et al.* (2020) who analyse the possibility of doing business through the arbitrage of electrical energy using Li Ion storage systems, and applying it to the Spanish electricity market. Using a mixed-integer linear programming model to optimize energy buying and selling, the authors conclude that if storage technology continues developing at the rate it has done in the last 20 years, the business could make sense in a short period of time.

Although Europe should be a single economic entity, there are currently various separate wholesale electricity markets. Despite the fact that most of them are strongly interconnected, prices show different levels and ranges. In addition, the financial risks and fiscal conditions of each country vary, which ultimately implies the existence of different discount rates. In summary, at least from a financial point of view, the European Union is composed of 27 different fiscal regimes, business risks, and financial and electricity markets.

There are numerous papers devoted to analysing the interest of battery arbitrage operations. Practically all of these studies analyse, for a specific country or market, the business opportunities given a storage technology and a required investment. Some of them also investigate the beneficial side effects arising from the deployment of storage systems (renewables and grid upgrading) –see for example Hou *et al*. (2017), Das *et al*. (2014) and Jannesar *et al*. (2018). Likewise, some authors propose the analysis of factors which can affect the business effectiveness (access tariffs, possibility of auctions to provide the service, or level of competition in the market), as is the case of Adebayo *et al*. (2018), Nasrolahpour *et al*. (2016) and Wu and Lin (2018).

Among the papers which mainly analyse different markets and storage technologies, we highlight the following; Walawalkar *et al*. (2007) study the sensitivity of the roundtrip efficiency on energy trading operations where the energy is stored in sodium sulphide batteries (SNa) and flywheels, determining the critical factors which affect the financial yield of each technology in the New York market (NYISO). Subsequently, Berrada *et al*. (2016) analyse the same market (NYISO), while expanding the technologies used to Compressed Air Energy (CAES) and Hydraulic Pumping System (PHS). They also increase the scope of the possible sources of value in the business, analysing the markets of daily, intraday and auxiliary services. An assessment of SNa battery plants and PHS is conducted by Kazempour *et al*. (2009) for the Alberta (Canada) wholesale market. In this case the authors apply a linear programming model which reveals that PHS technology offers a significant and clear alternative to SNa batteries in economic terms. In these three works, although the authors search for the optimal markets, technologies and participation strategies, the results are far from the required profitability thresholds.

For their part, Sioshansi *et al.* (2009) propose an optimization model for the Independent System Operator PJM. Considering a conservative roundtrip efficiency of 80% and a 12-hour storage capacity for large storage technologies, these authors estimate the economic value that arbitrage transmits to society by calculating the overall welfare and the surplus of consumers and producers in the day-ahead electricity wholesale market. As a main contribution, the authors present the breakeven cost of storage and describe what regulatory measures could be incorporated in order to make that business more attractive. In our opinion, although their model is meaningful, battery technology has improved significantly in the last ten years thereby justifying updating it to current values. From an international perspective, Connolly *et al*. (2011) analyse a number of strategies to optimize electricity arbitrage in the day-ahead market in eleven different countries, but without discriminating on the basis of fiscal and financial risk features; in general, the results obtained show that the most favourable strategy is the one that focuses on short-term trade (i.e., next 24 hours).

The contribution of Yucekaya (2013) consists of the application of a mixed integer optimization model based on Markov chains, with which it simulates arbitrage using exclusively CAES (Compressed Air Energy Storage) technology in the Turkish electricity market, and analyses the influence that each of the technical parameters of CAES has on business profitability. The same technology is being analysed by Das *et al*. (2014) for the Californian market, though using one-minute resolution demand data and allowing participation in the ancillary services market, thus making the operation even more profitable and enhancing it as wind power penetration increases.

There is more literature in this field; for example, we have found works that compare technical and financial pros and cons for different technologies, such as PHS, CAES, ZEBRA (nickel-sodium chloride battery), EDLC (ultracapacitors), lead-acid battery (LA), and Li-Ion batteries –see for instance Bradbury *et al*., 2014; Zakeri and Syri, 2014; and Terlouw *et al*., 2019. Other studies analyse different business models derived from the participation of storage agents in the electricity markets –He *et al*., 2011, Daggett *et al*., 2017; Berrada and Loudiyi, 2016; Adebayo *et al*., 2018; and Jannesar *et al*., 2018. For their part, Nasrolahpour *et al*. (2016) analyse how the more or less competitive character of the market affects the arbitrage business. Finally, in the case of China, since the electricity market is not competitive, Wu and Lin (2018) study the impact that the use of storage systems can have on the efficiency of the electrical system.

As far as we know, there are no studies which make a comparative analysis of several countries while also including in that analysis information on the characteristics of the wholesale electricity markets, financial risks and fiscal issues at country level. Our research assumes Li-Ion as the optimal storage technology, due to the great technological and economic advances registered in this type of battery in recent years, and assumes a competitive nature for the electricity wholesale market; consequently, the contribution of the storage player will not affect the equilibrium price (they are price taker). Likewise, we make an international comparison, adopting a common procedure to measure risk and tax levels of every country in the sample. These three aspects considered together give a distinctive character to our work within the existing literature in this field.

Our research raises some specific questions: 1) Will the different European wholesale electricity markets be equally interested in using battery arbitrage? If not, 2) What are the determinants of these differences? And finally, 3) In which country (and at what moment) would it be more profitable for a company to undertake the arbitrage business with storage? In order to address them, the following steps have been taken. First, the hourly prices of electricity distribution (daily market) in 24 European countries have been analysed, evaluating both their levels and daily gaps. Once we know how electricity markets perform, the discount rates of each country are estimated using the procedure applied by the Spanish regulator. The fact that there are different tax rates and risk premiums in each country implies that the values of the discount rates are consequently different. Parallel to the calculation of the discount rates, we use price data from the daily electricity market and BESS technological information to develop an optimization model which determines the best buy and sell strategies for different converter sizes (from 1 to 10 MW) given a battery of size 10 MWh, 5,000 cycles and 92% round trip efficiency. Finally, with the results of the purchase and sale of energy coming from the optimization model, and with the financial and fiscal information of each country, the financial viability of each BESS configuration, in each of the 24 countries considered, is analysed.

In addition to a broad international comparison, our study is novel in another sense. Once the profitability obtained from the arbitrage business has been determined for each country, we analyse the influence that each of the input variables (economic and technical) has on that profitability by estimating a two-level mixed regression model. Specifically, the model is estimated for both the net present value (NPV) and the internal rate of return (IRR) indicators. As we will see, the need to control for the existence of unobservable heterogeneity is important in these types of international comparisons. In our opinion, the conclusions and business implications provided by this study can be used by entrepreneurs to determine in which countries the business of battery arbitrage can be more or less profitable, as well as to know what factors affect the economic returns. This information can also be valuable for regulators and technology research centres.

After this introduction, the information used in this paper (at market, technical and financial level) is presented and comparatively analysed (by country) in section 2. Subsequently, section 3 develops the formulation of the mathematical model of optimal trading with storage. Section 4 calculates the financial results by country and investment configuration, using as input information the optimal cash-flows provided by the previous optimization model. Moreover, an econometric analysis is proposed in order to analyse the main determinants of the different financial returns observed, thus providing answers to the research questions proposed above in this introduction. Finally, section 5 concludes.

## 2. Data and materials.

### 2.1. Market information.

The transparency platform ENTSO-E collects data on generation, transmission and consumption for the pan-European market. This data collection has allowed us to observe the hourly wholesale electricity prices for the year 2019 in a sample of 24 European countries. Table 1 orders those countries from highest to lowest average hourly price (€/MWh). The table also shows the average daily differential between the maximum and minimum intraday hourly prices.

**Table 1.** Wholesale electricity prices in Europe (Source: ENTSO-E).

| Country | Hourly electricity prices | | | | Daily price differential | | | |
|---|---|---|---|---|---|---|---|---|
| | Mean price | Price Std. Dev. | Min | Max | Mean gap | Std. Dev. | Min | Max |
| Greece | 63.8 | 11.8 | 0.0 | 145.0 | 28.2 | 17.6 | 3.3 | 84.6 |
| Italiy | 52.2 | 12.7 | 1.0 | 113.1 | 29.8 | 9.2 | 9.2 | 58.5 |
| Serbia | 50.5 | 18.0 | 0.5 | 153.5 | 37.0 | 13.0 | 0.0 | 92.1 |
| Romania | 50.4 | 21.2 | 0.0 | 158.0 | 47.8 | 20.0 | 16.2 | 127.1 |
| Hungary | 50.4 | 18.8 | 0.0 | 138.8 | 42.0 | 17.2 | 14.1 | 103.0 |
| Ireland | 50.3 | 23.7 | -11.9 | 365.0 | 54.7 | 37.5 | 17.5 | 298.1 |
| UK | 50.2 | 23.8 | -11.9 | 365.0 | 54.6 | 37.6 | 0.0 | 298.1 |
| Poland | 49.4 | 17.5 | 1.2 | 114.0 | 19.5 | 9.6 | 1.5 | 63.2 |
| Croatia | 49.2 | 18.9 | -20.2 | 200.0 | 40.4 | 20.9 | 11.2 | 161.7 |
| Slovenia | 48.7 | 18.2 | -20.2 | 200.0 | 38.3 | 20.4 | 10.5 | 161.7 |
| Portugal | 47.9 | 10.8 | 0.0 | 74.7 | 16.3 | 7.6 | 2.2 | 50.4 |
| Spain | 47.7 | 10.9 | 0.0 | 74.7 | 16.9 | 8.3 | 2.2 | 55.2 |
| Latvia | 46.3 | 15.8 | 0.1 | 200.0 | 32.7 | 26.8 | 4.3 | 197.0 |
| Lithuania | 46.1 | 15.8 | 0.1 | 200.0 | 32.7 | 26.6 | 4.3 | 197.0 |
| Finland | 44.0 | 15.3 | 0.1 | 200.0 | 28.3 | 21.4 | 3.4 | 170.9 |
| Netherlands | 41.2 | 11.3 | -9.0 | 121.5 | 26.6 | 9.5 | 6.7 | 65.3 |
| Switzerland | 40.9 | 12.3 | -39.5 | 108.1 | 19.8 | 8.1 | 3.9 | 70.5 |
| Czech | 40.2 | 13.5 | -48.1 | 109.3 | 27.7 | 10.4 | 7.9 | 80.8 |
| Austria | 40.1 | 13.1 | -59.8 | 121.5 | 26.7 | 11.3 | 4.0 | 87.6 |
| France | 39.5 | 14.0 | -24.9 | 121.5 | 27.2 | 9.4 | 6.4 | 64.9 |
| Norway | 39.3 | 8.3 | 5.9 | 109.5 | 7.7 | 6.7 | 1.1 | 54.1 |
| Denmark | 38.5 | 13.2 | -48.3 | 109.5 | 26.3 | 13.8 | 4.1 | 105.2 |
| Sweden | 37.9 | 9.9 | 0.1 | 107.7 | 13.5 | 9.5 | 1.9 | 52.3 |
| Germany&Lux. | 37.7 | 15.5 | -90.0 | 121.5 | 30.1 | 15.6 | 6.7 | 117.3 |

Prices are relatively high in Southeast Europe and relatively low in Northern Europe, where the generation mix costs are lower. Note also that countries such as Spain and Portugal or the United Kingdom and Ireland have integrated markets, so they share the same prices for most of the year –if the interconnections were unlimited, the prices would be the same in both countries; only in those cases where the interconnections are saturated, different prices may appear. To better visualize the table, Figure 1 represents the average hourly price, its standard deviation and the average differential price for each country in the sample for the year 2019. As can be seen, the figure shows some positive correlation between the average price and the differential price. The United Kingdom and Ireland stand out for their high differential prices (even higher than their average prices) while Sweden and Norway stand out for their small differentials. It is also remarkable that Portugal and Spain show mid-level prices with relatively small standard deviations, which determines that their price daily gaps are relatively small in their price quartile.

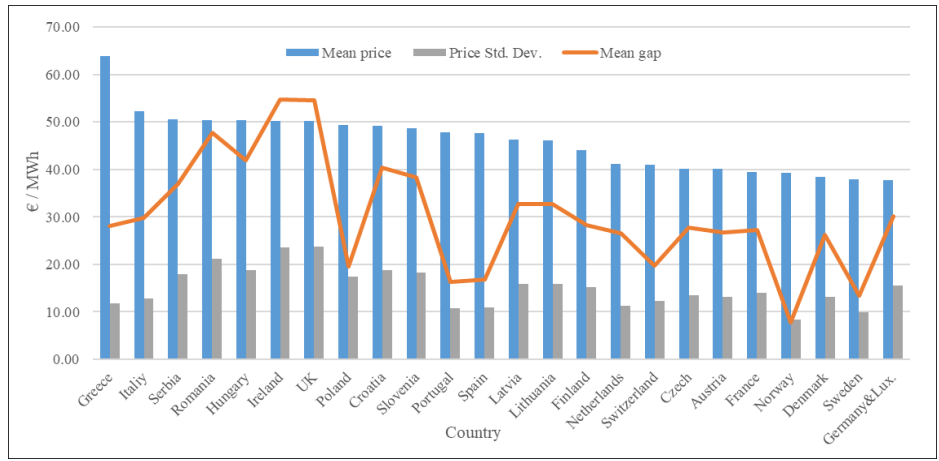

**Figure 1.** Electricity prices in Europe. Average levels and max.-min. daily differentials.

**2.2. Technical information.**

As stated in the introduction, Li-Ion batteries have been selected as the preferred storage technology due to their better level of performance and recent technological advancement. An analysis of the different storage technologies considered can be found in Arcos-Vargas *et al.* (2020). Table 2 shows the technical specifications of Li-Ion batteries, which have been obtained from various sources: Gomez-Expósito *et al.* (2017), Segui (2018), Bardo-Cáceres (2010), Hernández-Romero (2016), Battery University (2019), Jofemar Energy (2016), Vélez-Moreno (2012), CleanTechnica (2017) and IRENA (2017).

**Table 2.** Main features of Li-Ion batteries.

| Specific energy (Wh/kg) | 130-147 |
| Energy density (Wh/L) | 250-730 |
| Specific power (W/kg) | 250-340 |
| Nominal voltage (V) | 3.6 |
| Charge/discharge cycles | 5000 |
| Monthly self-discharge (%) | 3% |
| Round-trip efficiency (%) | 92% |
| CAPEX (€/kWh) | 100 |

The current performance of this type of storage technology has improved significantly in recent years. In particular, the cost has been reduced by an average of 20% per year during the period 2010-2019, and given the intensity of research focused on this field, it is expected that the trend of improvement will continue. If cost reduction continues at this rate, they would be halved in 3.5 years.

**2.3. Financial information.**

Although the European Union should be a common business area, the reality is still far off. The existence of different wholesale electricity markets, taxation regimes, and financial risks, motivated on several occasions by the decisions of their governments, means that different discount rates have to be applied depending on the country in which the installation is operated. The Weighted Average Cost of Capital (WACC) represents the minimum return that a company must earn, on an existing asset volume, to satisfy its creditors, owners, and other providers of capital, or they will invest in other more profitable activities (Fernandes, 2014). Subsequently, a reasonable differential must be added to this minimum return in order to obtain the discount rate (CNMC, 2018). The proposed model for evaluating the BESS arbitrage in different European countries is based on the following two hypotheses: 1) there is a unicity of capital goods, technology and capital markets, 2) there are country-specific characteristics in terms of taxation and business risks. From the first hypothesis, it is clear that the same technology and the same investment cost can be accessed in all European countries. Furthermore, given the principle of free capital flow, all markets will have the same market risk premium (*MRP*) and the same *β* coefficient for the business associated with the operation in the electricity markets. The second assumption implies different fiscal systems and, therefore, different profit taxes ($T_i$).

To estimate the discount rate for each European country $i$ ($DR_i$), the first step consists in calculating the $WACC$ value for each country –for those companies which participate in their respective wholesale electricity markets. For this purpose, the well-known formula of the $WACC$ is used:

$$WACC_i = \frac{E}{E+D} r_{E_i} + \frac{D}{E+D} r_D (1 - T_i) \qquad (1)$$

Where $E/(E+D)$ and $D/(E+D)$ are respectively the equity and debt fractions of total capital employed, $r_D$ is the cost of debt, and $T_i$ and $r_{E_i}$ are the profit tax and the cost of equity of each country $i$ respectively. To obtain the cost of equity ($r_{E_i}$), the CAPM (Capital Asset Pricing Model) method is applied, as it is the most widely used in the financial context:

$$r_{E_i} = RRF_i + \beta_{elec}\, PMR \qquad (2)$$

Substituting (2) in (1), we obtain:

$$WACC_i = \frac{E}{E+D}(RRF_i + \beta_{elec}\, PMR) + \frac{D}{E+D} r_D (1 - T_i) \qquad (3)$$

where $RRF_i$ represents the risk-free rate of return in country $i$, $\beta_{elec}$ is the beta coefficient of European companies in the electricity market, and $PMR$ is the Premium Market Risk.

Two sets of parameters can be differentiated in equation (3): those that present a common value for all the countries of the sample $\{E/(E+D), D/(E+D), \beta_{elec}, PMR, r_D\}$, and those whose values will depend on the country in which the activity is performed $\{RRF_i, T_i\}$. In order to assign values to the first group of parameters (common information), we take representative values from a group of European listed companies which are mainly active in the wholesale electricity market. These values are taken from the Spanish regulator (CNMC, 2018, BOE, 2019), which, in turn, uses data from Bloomberg (2019) and Dimson *et al.* (2018). Specifically, the estimated values are: $\{\frac{E}{E+D} = 0.45, \frac{D}{E+D} = 0.55, \beta_{elec} = 0.77, PMR = 4.75\%, r_D = 4.49\%\}$.

For the second set of variables, the country-varying variables $\{RRF_i, T_i\}$, the risk-free rates for each country ($RRF_i$) have been calculated by subtracting the 10-year sovereign bond yields of each country from that of Germany (OCDE 2020, and Trading economics, 2020), while the tax profit ($T_i$) has been obtained from the "Deloitte International Tax Source Report" (Deloitte, 2020). Finally, for the determination of the Discount Rate ($DR_i$), we apply the procedure used by the Spanish CNMC when calculating the rate of remuneration for regulated activities. The only parameter which has not been specified so far is the spread for additional risks, which, for the sake of simplicity, and since the peculiarities of each country are not known in depth, is maintained for all countries at the 0.5% proposed by the CNMC for the Spanish electricity distribution. Equation (4) represents the formula used to obtain the Discount rates:

$$DR_i = \frac{WACC_i - T_i + Spread}{1 - T_i} \qquad (4)$$

Table 3 shows all the variables that change depending on the country considered –countries are ranked from highest to lowest $DR_i$.

**Table 3.** Financial variables by country.

| Country | $T_i$ | $RRF_i$ | $r_{Ei}$ | $WACC_i$ | $DR_i$ |
|---|---|---|---|---|---|
| Greece | 28.0% | 2.10% | 5.76% | 4.37% | 7.07% |
| Croatia | 18.0% | 2.75% | 6.41% | 4.91% | 6.99% |
| Italy | 24.0% | 1.79% | 5.45% | 4.33% | 6.69% |
| Romania | 16.0% | 2.25% | 5.91% | 4.73% | 6.63% |
| Spain | 30.0% | 0.81% | 4.47% | 3.74% | 6.34% |
| Serbia | 15.0% | 1.75% | 5.41% | 4.53% | 6.33% |
| Czech | 19.0% | 1.25% | 4.91% | 4.21% | 6.20% |
| Portugal | 21.0% | 0.96% | 4.62% | 4.03% | 6.10% |
| Slovenia | 19.0% | 1.01% | 4.67% | 4.10% | 6.06% |
| Norway | 22.0% | 0.74% | 4.40% | 3.91% | 6.01% |
| Latvia | 20.0% | 0.80% | 4.46% | 3.98% | 5.98% |
| France | 33.3% | 0.03% | 3.69% | 3.31% | 5.96% |
| Poland | 19.0% | 0.75% | 4.41% | 3.98% | 5.92% |
| Hungary | 9.0% | 1.15% | 4.81% | 4.41% | 5.85% |
| Lithuania | 15.0% | 0.75% | 4.41% | 4.08% | 5.80% |
| UK | 19.0% | 0.30% | 3.96% | 3.78% | 5.67% |
| Austria | 25.0% | -0.01% | 3.65% | 3.49% | 5.66% |
| Netherlands | 25.0% | -0.17% | 3.49% | 3.42% | 5.56% |
| Sweden | 21.4% | -0.13% | 3.53% | 3.53% | 5.49% |
| Ireland | 12.5% | 0.16% | 3.82% | 3.88% | 5.43% |
| Finland | 20.0% | -0.23% | 3.43% | 3.52% | 5.40% |
| Denmark | 22.0% | -0.44% | 3.22% | 3.37% | 5.33% |
| Germany&Lux. | 15.0% | -0.47% | 3.19% | 3.53% | 5.16% |
| Switzerland | 8.5% | -0.49% | 3.17% | 3.68% | 5.03% |

The discount rate and the profit tax will be used in the following sections to carry out our financial analysis. Figure 2 depicts these two variables. As can be observed, southern European countries have high discount rates compared to central and northern European countries. Moreover, there is a certain positive correlation between the discount rate (left axis) and the profit tax (right axis), with Spain and France standing out for having relatively high tax rates.

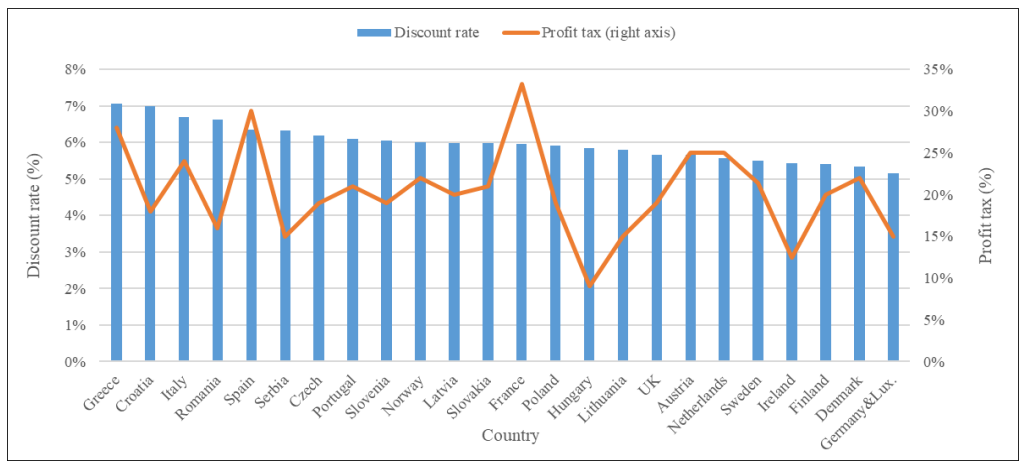

**Figure 2.** Discount rates and profit taxes by country.

**3. Trading model formulation.**

This section describes the optimization trading model designed for determining the optimal purchase and sale strategy for a given battery capacity. The specific battery size does not influence the optimal results, since the operation policy is conditioned by the ratio 'battery capacity/converter transformation'. The objective of the proposed trading model consists of the maximization of cash-flows obtained from energy purchase and sale operations during certain planning horizons. Later, we will explain how the proposed optimization framework deals with multi-year planning intervals. As its main input, the model considers hour-by-hour electricity prices. Figure 3 shows a plausible buying and selling schema, and it is used to explain the model characteristics more clearly. At the top of the Figure, series A depicts a possible hourly pattern of the electricity price, each interval corresponding to one hour. A rational policy would consist of buying energy when market prices are low and selling it when prices are high. The middle part of the Figure illustrates a feasible strategy. Filled and dashed rectangles represent purchasing and sales decisions respectively. The height of each rectangle corresponds to the amount of energy charged or uncharged. Numbers (1) and (2) are related to charging operations whereas numbers (3) and (4) depict discharging ones. The line C represents the battery charge level. As shown, a purchase operation turns into an increment in the battery charge level. The dashed line B measures the maximum battery capacity. As illustrated, each time a new charge/discharge cycle takes place, the maximum battery capacity must be decreased, representing in this way the battery deterioration (depreciation in economic terms) with use. Next we explain in detail the model parameters, variables and the set of constraints defining the set of feasible solutions.

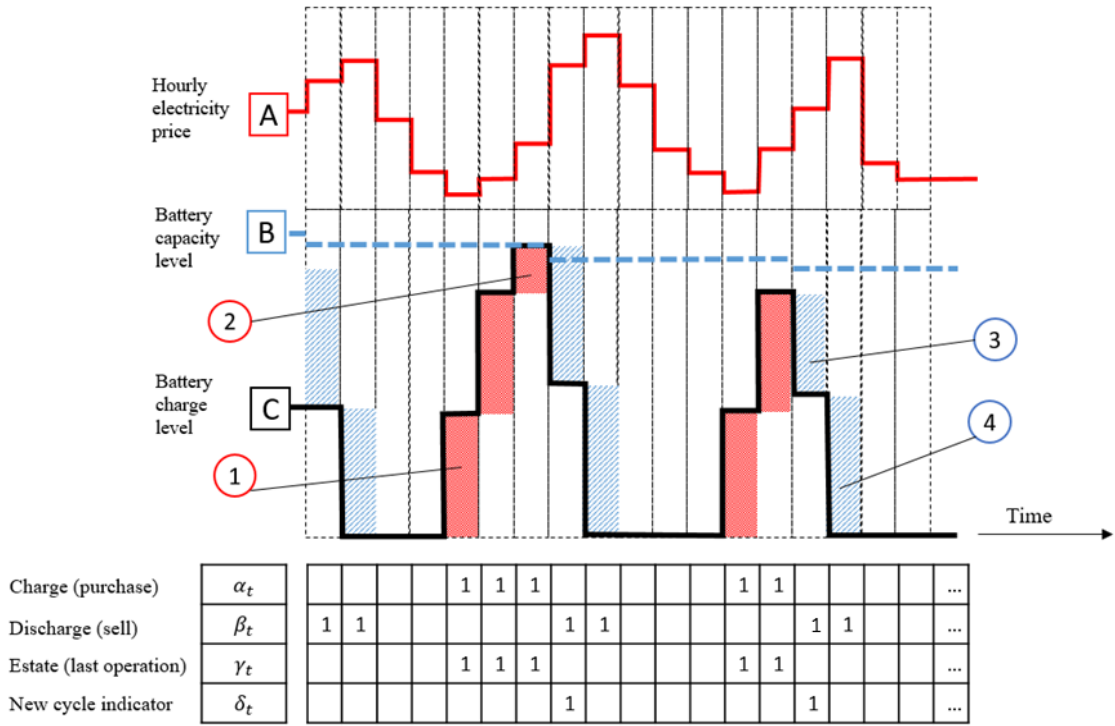

**Figure 3.** Illustration of buying and selling operations.

*Parameters*

$T$     Set of periods (Hours) defining the planning horizon.

$\omega_t$     Market electricity price at hour $t \in T$.

$\psi_o$     A parameter used to define the initial battery capacity (we will consider a battery of 10 MWh in our experiments).

$\varphi$     Deterioration of the battery capacity per cycle. We consider that after 5000 cycles the initial battery capacity decreases up to a 20% of the initial one.

$\psi_{max}$     Conversion rate of the converter.

$\lambda$     A parameter that defines the loss of energy in charging and discharging conversion processes. This parameter is measured as a percentage of the energy purchased or sale.

As previously defined, the model parameters are determined by the battery and converter technologies. Specifically, we consider: the round-trip efficiency in the charge\discharge process ($\lambda$), the initial capacity of battery ($\psi_o$), the conversion capacity ($\psi_{max}$) and the useful life of the battery (given by the deterioration factor $\varphi$, i.e., the number of charging\discharging cycles until the effective battery capacity decreases to 20% of the initial one). It is assumed that the battery capacity diminishes in a linear way with the number of charging\discharging cycles and that the operation and maintenance battery costs are 1% of the equipment value.

*Variables.*
  a) *Binary variables to simulate the behaviour depicted in Figure 3*

$\alpha_t$   Binary variable. It takes value 1 if the battery is charged at period (hour) $t \in T$, 0 otherwise. As illustrated in Figure 3, these variables are activated each time a new charging process is performed, for instance, numbers (1) and (2). Note that in the table at the bottom of Figure 3, for clarity, 0 values are not represented.

$\beta_t$   Binary variable. It takes value 1 if the battery is discharged at period $t \in T$, 0 otherwise. They correspond, for instance, to numbers (3) and (4) in Figure 3.

$\delta_t$   Binary variable. It takes value 1 if the battery has finished a charge\discharge cycle at period $t \in T$, 0 otherwise. For illustration purposes, these variables are represented in the last row of the table at the bottom of Figure 3. They are in charge of measuring the battery deterioration with use.

$\gamma_t$   Binary variable. It takes value 1 if the last performed operation prior to period $t \in T$ was a charge and 0 if it was a discharge. These variables will be used as auxiliary variables to enforce cycle occurrences.

  b) *Real variables to measure amounts of stored energy, purchases and sales.*

$P_t$   Represents the amount of electricity purchased and stored in the battery during period $t \in T$. Graphically, $P_t$ are represented by the height of the rectangles (1) and (2) in Figure 3. The purchase decision at period $t$ is conditioned by $\alpha_t$.

$S_t$   Measures the amount of energy discharged from the battery during period $t \in T$. This energy is transformed into electricity and sold at the same period. Variables $S_t$ are depicted as the height of the rectangles (3) and (4) in Figure 3. The sale decision at period $t$ is conditioned by $\beta_t$.

$C_t$   Integer variable that measures the cumulative number of charging\discharging cycles that has been done just until period $t \in T$.

$K_t$   Represents the remaining capacity of the battery at the end of period $t \in T$. It is represented by line B in Figure 3.

$L$   Measures the level of energy in the battery at the end of period $t \in T$. It is represented by line C in Figure 3.

The mathematical formulation is as follows:

$$Max \sum_{t \in T} \omega_t (S_t - P_t) \quad (5)$$

$$P_t \leq (K_{t-1} - L_{t-1})/(1 - \lambda), \quad t \in T \quad (6)$$

$$S_t \leq L_{t-1} (1 - \lambda), \quad t \in T \quad (7)$$

$$P_t \leq \psi_{max}\, \alpha_t, \quad t \in T \quad (8)$$

$$S_t \leq \psi_{max}\, \beta_t, \quad t \in T \quad (9)$$

$$\alpha_t + \beta_t \leq 1, \quad t \in T \quad (10)$$

$$L_{t-1} + P_t (1 - \lambda) - S_t/(1 - \lambda) = L_t, \quad t \in T \quad (11)$$

$$K_t = K_{t-1} - \varphi \cdot \psi_o \cdot \alpha_{t-1}, \quad t \in T \quad (12)$$

$$\gamma_1 = 0 \quad (13)$$

$$\gamma_{t-1} - \gamma_t \leq \delta_t, \quad t \in T \setminus \{1\} \quad (14)$$

$$\alpha_t \leq \gamma_t, \quad t \in T \quad (15)$$

$$\gamma_t \leq 1 - \beta_t \quad t \in T \quad (16)$$

$$\gamma_{t-1} - \beta_t - \alpha_t \leq \gamma_t \quad t \in T \setminus \{1\} \quad (17)$$

$$\gamma_{t-1} + \beta_t + \alpha_t \geq \gamma_t \quad t \in T \setminus \{1\} \quad (18)$$

$$C_o = 0 \tag{19}$$

$$C_t = C_{t-1} + \delta_t \quad t \in T\setminus\{1\} \tag{20}$$

$$\alpha_t, \beta_t, \gamma_t, \delta_t \in \{0,1\}, \quad t \in T$$

$$P_t, S_t, K_t, L_t \geq 0, t \in T$$

$$C_t \in \mathbb{N}^+, t \in T \tag{21}$$

The objective function (5) maximizes the cash-flows obtained from the battery operation during the planning horizon. At each period $t$, the amount of electricity purchased, when $\alpha_t = 1$, which is represented by $P_t$, must be less than or equal to the available battery capacity, as stated in constraints set (6), which is unknown and measurable by the term $(K_{t-1} - L_{t-1})/(1-\lambda)$. Also, the amount of energy will be constrained by the conversion capacity (8). These two constraints are represented by numbers (6) and (5) respectively in Figure 3. As previously mentioned, $\lambda$ measures the loss of energy as a consequence of the conversion process which takes place at the converter. Then, $(1-\lambda)$ stands for the roundtrip efficiency of the electricity conversion process. Concerning the discharge process, constraints (7) and (9) limit the electricity that can be discharged and sold at period $t$, $S_t$. In the first case, the discharged energy must be less than or equal to the level of charge of the battery plus the energy lost in the conversion process. Then, at each period $t$, the maximum possible discharge is $L_{t-1}(1-\lambda)$. In the second case, electricity sales are also bounded by the converter capacity trough constraint set (9). These constraints incorporate binary variables $\alpha_t$ and $\beta_t$ into the right-hand side, so that, when they are 0, $P_t$ and $S_t$ are also 0. Since charging and discharging operations cannot be performed in the same period, constraints set (10) are incorporated into the model.

The temporal evolution of the battery level ($L_t$) –see line C in Figure 3–, which depends on whether a purchase or sale of electricity has been done at period $t-1$, is modelled by constraints set (11). At each period $t$, the battery level is measured by adding or subtracting the charge ($P_t$) or discharge ($S_t$) of energy to the previous battery level $L_{t-1}$. The battery behavior also takes into account the decrement of the battery capacity with time, represented by line B in Figure 3, which is due to the increment in the number of charging\discharging cycles. This deterioration is modelled by constraints set (12). Note that binary variables $\delta_t$ control the occurrence of a new charging/discharging cycle.

In order to properly model the behaviour of variables $\delta_t$ a new set of binary variables is needed. To this end, binary step state variables $\gamma_t$ are introduced. They represent the last battery performed operation before period t. Suppose, for instance, that the battery was charged at period $t = 4$ and no other battery operation occurs until period $t = 7$, where a discharge operation takes place. In this example $\alpha_4$ must be equal to 1(a charge), then $\gamma_4 = \gamma_5 = \gamma_6 = 1$ (no operation has been done from periods 5 to 4) and finally $\gamma_7 = 1$ (since a discharge has taken place). At this point, $\delta_7$ must be equal 1, since a new charging\discharging cycle must be computed and the battery capacity must be decreased in $\varphi$, which is done by constraint set (20). Constraints set (13)-(18) are used to model this behavior in a general way. In the initial period, the step state variable is set to zero (13). For the rest of the periods, as stated in (14), if $\gamma_{t-1}$ is 1 (the last operation before $t$ was a charge one) and $\gamma_t = 0$ (the battery has been discharged at period $t$), the variable $\delta_t$ must be set to 1. The step state variables are also related to the binary charging\discharging variables, $\alpha_t$ and $\beta_t$ respectively. On one hand, if $\alpha_t = 1$, a charging operation has been done at period $t$, then $\gamma_t$ must be one, as enforced by constraints (15). On the other hand, if a discharge operation has been done at period $t$, $\beta_t = 1$, the battery step state variable $\gamma_t$ must be set to zero, as imposed by constraints (16). When the last performed battery operation at time $t-1$ was a charge ($\gamma_{t-1}=1$) and no charge or discharge has been done at time t, $\gamma_t$ must be 1, according to constraints set (17). Finally, when at time $t-1$ the last performed battery operation was a discharge ($\gamma_{t-1} = 0$) and no charge or discharge operation has been done at period $t$ ($\alpha_t = \beta_t = 0$), the variable $\gamma_t$ must be equal to zero, as imposed by constraints (18). Constraints (21) define the domain of variables. The next table summarizes these constraints:

**Table 4.** Relationships among binary decisions variables.

| $\gamma_{t-1}$ | $\alpha_t$ | $\beta_t$ | | $\gamma_t$ | Constraints set |
|---|---|---|---|---|---|
| 0 | 0 | 0 | → | 0 | (18) |
| 0 | 1 | 0 | → | 1 | (15) |
| 0 | 0 | 1 | → | 0 | (16) |
| 1 | 0 | 0 | → | 1 | (17) |
| 1 | 1 | 0 | → | 1 | (15) |
| 1 | 0 | 1 | → | 0 | (16) |

Constraints set (21) are used to specify the domain of the model variables.

*- Optimization framework implementation issues.*

First, since the purchase and sale of electricity per hour depends on the converter conversion rate, the optimal trading strategy must be obtained by varying the converter size, using the hourly prices of electricity as an input. Second, at the first glance the length of the planning horizon is unknown since it represents the number of operation periods (hours) until the battery capacity reaches a low enough residual value (in our case 20% of the initial battery capacity) which cannot be determined before running the optimization model –note that the battery deterioration is a consequence of the specific policy followed. Since we are dealing with hourly prices, there are 8,670 time periods per year (24 hours per day times 365 days per year). For only one year the model size rises 100,765 constraints, 30,654 continuous variables and 35,033 binary variables. Moreover, the length of $T$ will be an unknown multiple of 8,670 time periods.

To reasonably resolve these shortcomings, we apply a sequential strategy in order to solve the problem; this is depicted in Figure 4. As shown in the Figure, we opt for solving the problem in a sequential way, year by year, beginning each year of the battery operation by using the residual battery capacity obtained after solving the previous year (the upper index in the initial battery levels corresponds to the year of operation). This process is repeated a certain number of times (depending on the converter size considered) until the battery capacity is exhausted.

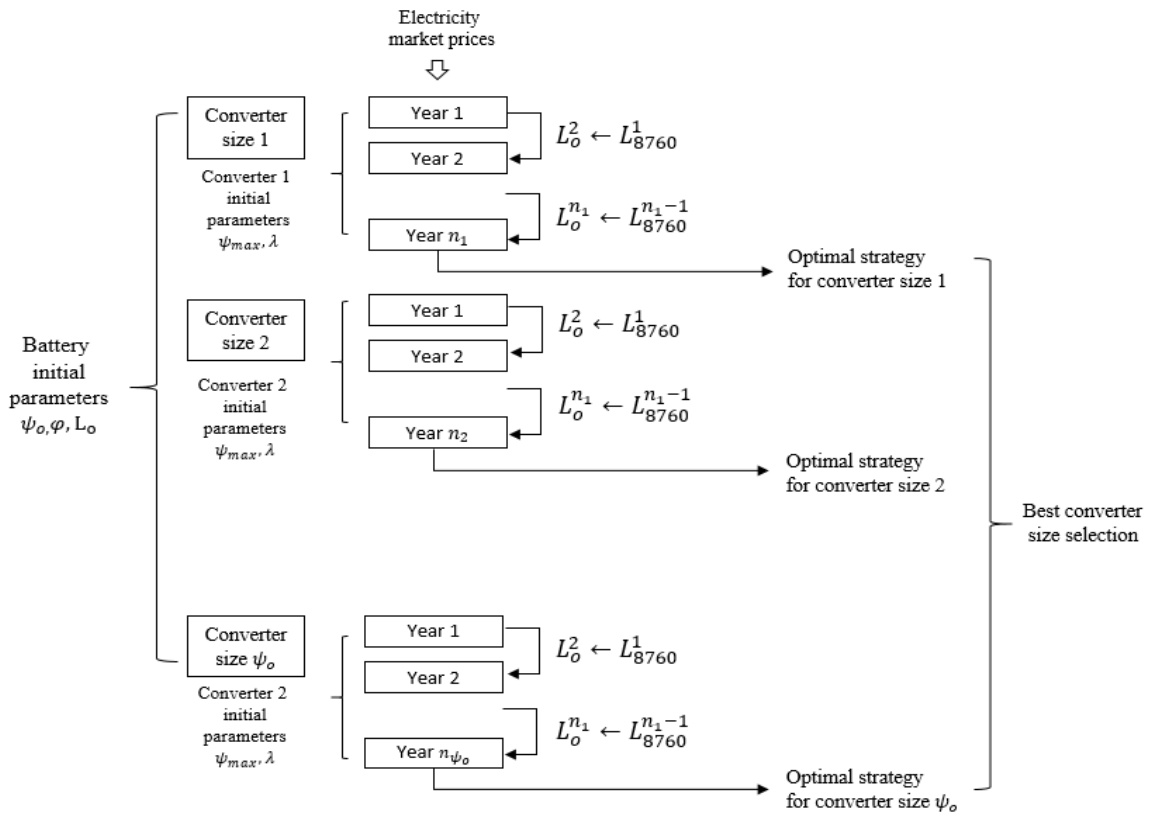

**Figure 4.** Optimization framework structure converter.

Without loss of generality, we consider an initial battery capacity of $\psi_o$ =10 MWh. The specific battery size considered to solve the problem does not affect the results since they are completely scalable. Which is really important is the rate 'battery capacity/converter size'. Consequently, we solve the optimization model by varying the converter size from 1 to $\psi_o = 10\ MWh$, covering rates from 10/1 to 1. For each converter size, we repeat the optimization year by year, until the maximum battery deterioration is reached. We are assuming that, in any case, the battery size is not enough to influence the market's operation.

## 4. Financial results and econometric analysis.

In this section, we analyse the financial aspects of the optimal buying/selling activity determined in the previous section for the 10 MWh battery with 92% of round-trip efficiency and an expected life of 5000 cycles. The following figure (Figure 5) shows the hours when, during the first operating year and for the different countries, the battery is either buying electricity, selling it, or staying inactive. To simplify the figure, we represent only four of the ten converter sizes considered (1 MW, 4 MW, 7 MW, and 10 MW) and rank the countries in each plot according to the hours of battery inactivity. Several aspects stand out in the figure. On the one hand, the higher the converter, the lower the number of hours the battery is active (buying or selling). On the other hand, the total hours buying and the total hours selling are quite similar in each country, which would indicate that the battery charges or discharges using all the capacity of the converter. Finally, different behaviour is observed between countries and converters. For example, for the 4 MW converter, the most active countries (France, Netherlands, Ireland and the UK) spend approximately 2000 hours buying and 2000 hours selling, while the rest of the year (4760 hours) the battery remains inactive; these values contrast with those observed in countries like Norway or Sweden, where the battery is approximately 1000 hours buying and 1000 hours selling during the first year of operation. Note that these operational differences between countries imply that the battery lasts longer in some countries than others. Therefore, in each country, the business will last until the battery runs out or until it begins to generate negative cash flows (if this occurs); in this last case the residual value of the equipment (battery and inverter) would be recovered.

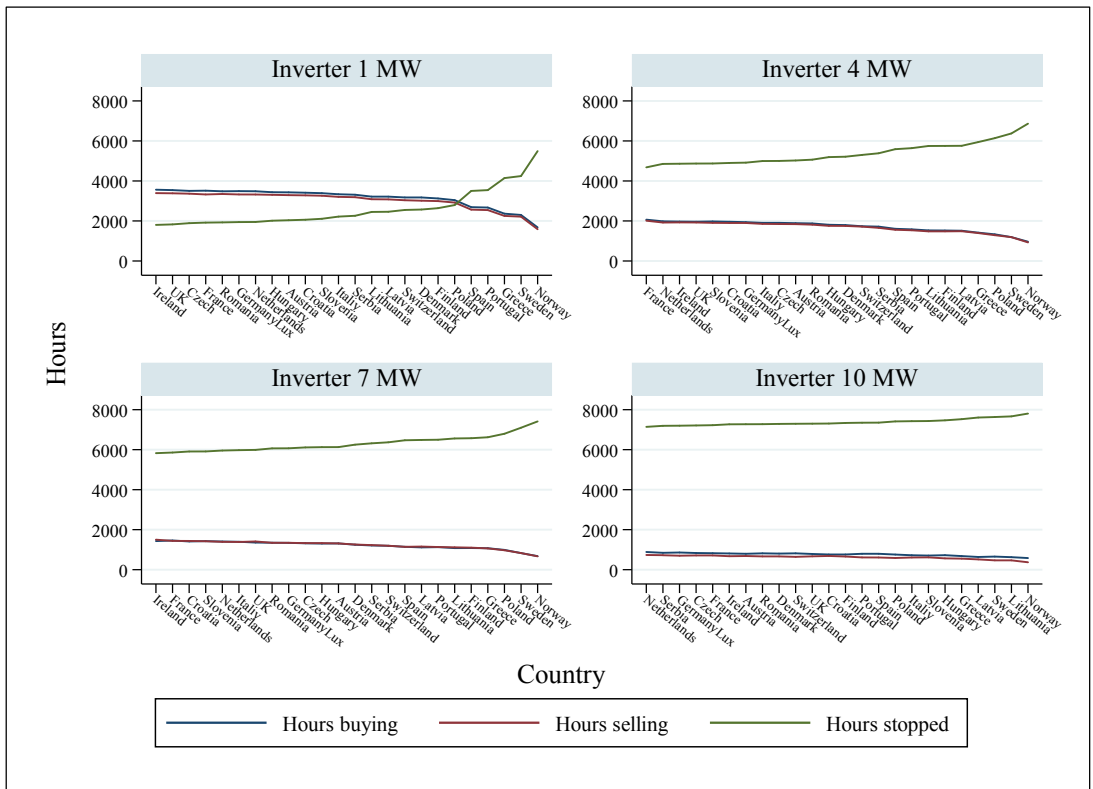

**Figure 5.** First year of trading activity by country and converter.

To further investigate the heterogeneity among countries, we represent in Figure 6 the two countries with the greatest difference in trading strategy when the converter size is 7 MW, Ireland and Norway. As can be seen, prices in Ireland are on average somewhat higher than in Norway and show a much greater volatility. These two facts help explain one of the results of our financial analysis: the BESS arbitrage is more profitable in Ireland than in Norway.

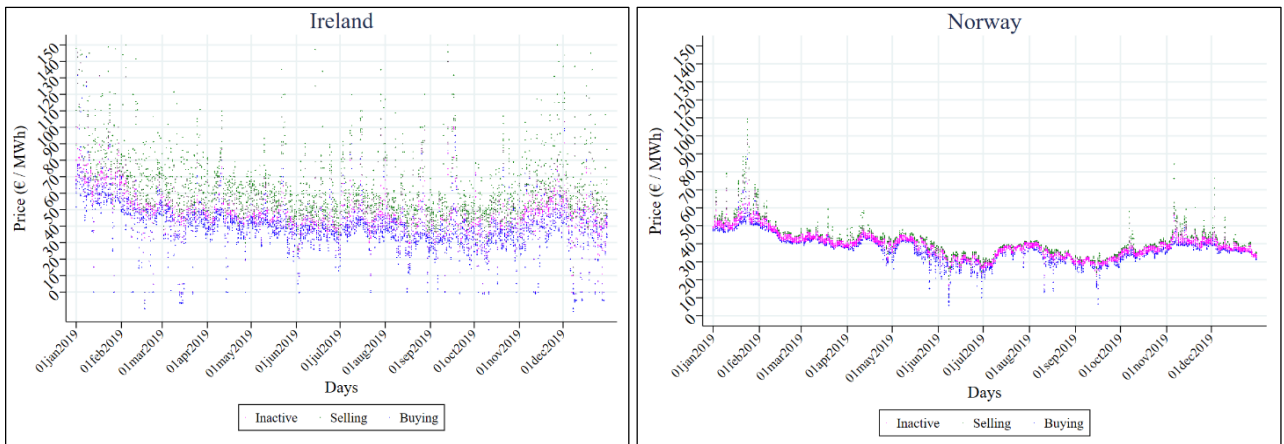

**Figure 6.** First year of BESS arbitrage in Ireland and Norway (7 MW converter).

Our optimization model guarantees optimal arbitrage with a 10 MWh battery for each country and converter (from 1 MW to 10 MW), given the hourly electricity prices in each country (during the year 2019) and the costs of the battery (100,000 €/MWh) and the converter (30,000 €/MW). With this optimal trading information, we can generate annual cash-flows by country and converter, and calculate for each case two financial indicators, the Net Present Value (NPV) and the Internal Rate of Return (IRR). We restrict the coming financial and econometric analysis to the most profitable converter sizes in each country according to those financial indicators –as represented by Figure 7. The respective converter sizes which maximize NPV and IRR indicators do not have to coincide within each country or to be the same across countries –as can be seen in the table attached to Figure 7. All countries move in negative values of both indicators (the business is not profitable yet), and there is a high positive correlation between the two indicators. Another interesting fact is that the best converters in terms of NPV do not exceed 6 MW of capacity, while in IRR terms they are usually equal to or greater than 6 MW, with some exceptions such as Sweden or Norway.

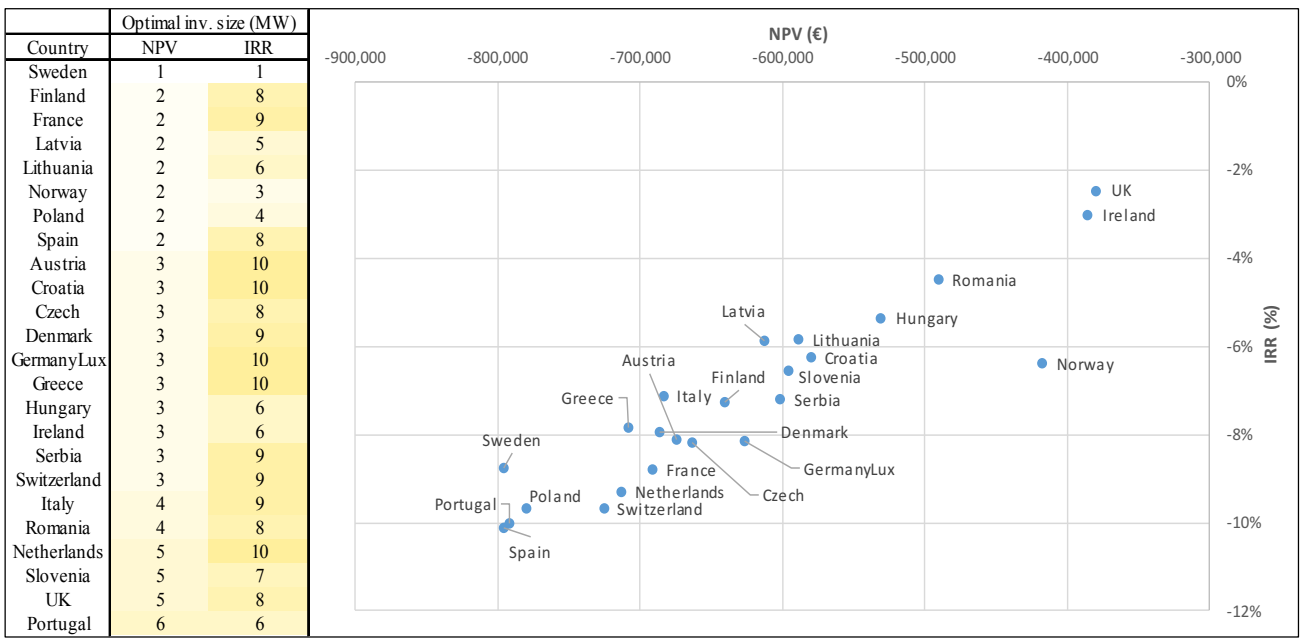

**Figure 7.** Net Present Value and Internal Rate of Return by country.

The economic and technological information dealt with in this article allows us to study the main determinants of the NPV and IRR indicators. For this purpose, we propose the estimation of two mixed linear regression models, one for each indicator. The linear mixed or multilevel regression model can be seen as a generalization of the linear regression model which allows the inclusion of random deviations other than those associated with the overall error term of the model –for multilevel analysis details, see for example Cameron and Trivedi (2005) and Goldstein (2011). This model has been implemented in several fields of research –education, medicine, labour market, etc.– but its presence in studies on electricity arbitrage is scarce at present. As far as we know, there are no previous studies that make use of in-sample predictions, in a multilevel residuals scheme, to compare European countries with different electricity prices and different fiscal and financial conditions.

After trying several specifications, we propose three two-level models (called *I*, *II* and *III* respectively) which can be respectively estimated for the dependent variables $NPV_{ic}$ and $IRR_{ic}$, where the subscript 'i' represents a particular investment (or BESS configuration) and 'c' represents the country where it takes place.

**Level 1 model:**

$$NPV_{ic}(or\ IRR_{ic}) = \alpha_{0c} + \delta_{1c}I^{ic}_{1MW} + \delta_{2c}I^{ic}_{2MW} + \cdots + \delta_{10c}I^{ic}_{10MW} + \beta_{1c}C^{ic}_{0.5} + \beta_{2c}C^{ic}_{1} + \beta_{3c}C^{ic}_{1.5} + \gamma_1 TAX_c$$
$$+ \gamma_2 MeanPrice_c + \gamma_3 SDPrice_c + \gamma_4 MeanGap_c + \gamma_5 SDGap_c + \alpha_1 DUR_{ic} + \alpha_2 DUR^2_{ic}$$
$$+ \alpha_3 DUR^3_{ic} + \varepsilon_{ic}$$

**Level 2 model:**

$$\alpha_{0c} = \tau_{00} + u_{0c}\ (random\ intercept);$$
$$\delta_{kc} = \theta_{k0} + w_{kc} \quad k = 1,\ldots,10\ (random\ slopes\ of\ converter\ cost\ dummy\ variables) \quad (22)$$
$$\beta_{jc} = \tau_{j0} + v_{jc} \quad j = 1,2,3\ (random\ slopes\ of\ battery\ cost\ dummy\ variables)$$

$$\text{Model } I: \varepsilon_{ic}\ iid \sim N(0,\sigma^2_\varepsilon), u_{0c}\ iid \sim N(0,\sigma^2_{u_0}), v_{kc} = 0, w_{jc} = 0$$
$$\text{Model } II: \varepsilon_{ic}\ iid \sim N(0,\sigma^2_\varepsilon), u_{0c}\ iid \sim N(0,\sigma^2_{u_0}), v_{jc}\ iid \sim N(0,\sigma^2_{v_k}), w_{kc} = 0$$
$$\text{Model } III: \varepsilon_{ic}\ iid \sim N(0,\sigma^2_\varepsilon), u_{0c}\ iid \sim N(0,\sigma^2_{u_0}), v_{jc}\ iid \sim N(0,\sigma^2_{v_j}), w_{kc}\ iid \sim N(0,\sigma^2_{w_k})$$
$$cov(\varepsilon_{ic},u_{0c}) = 0, cov(\varepsilon_{ic},v_{jc}) = 0, cov(\varepsilon_{ic},w_{kc}) = 0, cov(u_{0c},v_{jc}) = 0, cov(u_{0c},w_{kc}) = 0, cov(v_{jc},w_{kc}) = 0$$

The fixed part of the model has the following explanatory variables: the global average of the financial indicator ($NPV_{ic}$ or $IRR_{ic}$, depending on the estimate) for all the countries ($\tau_{00}$); the dummy variables $\{I^{ic}_{1MW}, I^{ic}_{2MW}, \ldots, I^{ic}_{10MW}\}$ which control for the cost of the different converters according to their size, € 30,000 being the cost of an additional MW of converter capacity (the 4 MW converter is the reference unit in the estimation); the dummy variables $\{C^{ic}_{0.5}, C^{ic}_{1}, C^{ic}_{1.5}\}$ which respectively control for three possible prices {0.5 M€, 1 M€, 1.5 M€} for the 10 MWh battery (2 M€ is the reference cost in the estimated model); and the rest of the explanatory variables which collect idiosyncratic information from each country, such as the profit tax ($TAX_c$), the mean ($MeanPrice_c$) and standard deviation ($SDPrice_c$) of hourly electricity prices in 2019, the mean ($MeanGap_c$) and standard deviation ($SDGap_c$) of the daily differential between the maximum and minimum hourly electricity prices in that year, and the duration of each investment in each country ($DUR_{ic}$), which enters the model with a cubic format –the discount rate is not statistically significant in the estimated models, possibly due to its low variance.

The random part of the model has the following purely random effects:

$u_{0c}$: Specificity (level 2 random intercept) of every particular country.

$v_{jc}$ ($j = 1,2,3$): Level 2 cross-effect that every particular country has on the slopes of the dummy variables $\{C^{ic}_{0.5}, C^{ic}_{1}, C^{ic}_{1.5}\}$ which control for the battery cost –our model suggests that unobserved heterogeneity at country level determines that the effect of the battery cost on the financial indicators depends on the particular country in which we are.

$w_{kc}$ ($k = 1,\ldots,10$): Level 2 cross-effect that every particular country has on the slopes of the dummy variables $\{I^{ic}_{1MW}, I^{ic}_{2MW}, \ldots, I^{ic}_{10MW}\}$ which control for the converter cost –analogously, our model supports the idea that the effect of the converter cost on the financial indicators may depend on the particular country in which we are.

$\varepsilon_{ic}$: Overall or level 1 error term.

The most complete estimate (model *III*) assumes that *purely random effects* affect the intercept of the model and the slopes of the dummy variables which control for the battery and converter costs. Dummy fixed effects and random effects ultimately determine that the expected value of the NPV (or IRR) of a specific BESS configuration in a specific country can move away from the global average of all countries.

Our mixed models have to be estimated by using maximum likelihood techniques (or by Bayes methods) since they have composite error terms whose variance is partitioned into the between-country variance components (the variances of the level 2 residuals) and the between-investment variance component (the variance of the level 1 residuals). Table 5 shows the results of estimating the models *I, II* and *III* using as dependent variables NPV and IRR. The likelihood-ratio $\chi^2$ test of null hypothesis of *no difference in fit between nested models* favours these specifications against other nested alternatives such as the linear regression model or the linear regression model with country fixed effects. Furthermore, among these three models, model *III* is the preferred one; this model is the one that shows a higher log likelihood function and a lower AIC and BIC criteria, so it will be our reference model.

Regarding the variables which refer to each investment opportunity within each country, we highlight the following evidence. Controlling for the other factors, the best converters in terms of NPV are those with a size between 3 MW and 5 MW, while in terms of IRR the two largest converters stand out (9 and 10 MW). For example, operating with the 10 MW converter (compared to that of 4 MW) reduces NPV by approximately 88,392.1 €, while increasing IRR by 0.68 percentage points.

The battery cost is a suitable indicator to simulate technological improvement in BESS. As battery cost is expected to fall in the next years, the model has considered four possible costs for the 10 MWh battery: 200,000 €/MWh (the reference price in the estimations), 150,000 €/MWh, 100,000 €/MWh and 50,000 €/MWh. The estimations show that most of the price decline is transferred to the NPV indicator, while the IRR indicator improves significantly with such a fall; for example, according to model *III*, if the battery is worth 100,000 €/MWh (instead of 200,000 €/MWh), the NPV indicator improves about 898,523.1 € and the IRR indicator does so at 2.47 percentage points.

The rest of the regressors of the fixed part of the model refer to idiosyncratic characteristics of each country (level 2 explanatory variables). The profit tax (expressed as a percentage) negatively, although moderately, affects the profitability indicators; that is, a unit increase in the tax rate reduces the NPV by around 3,221.5 € and the IRR by 0.07 percentage points. The variables which control the behaviour of hourly electricity prices show that the mean and standard deviation of the daily price differential (variables MeanGap and SDGap respectively) positively affect the business profitability. The coefficients of the mean and standard deviation of electricity price (variables MeanPrice and SDPrice respectively) turn out not to be significant in the model, except in the case of model *I*-IRR, where the mean price shows a significant positive effect of 0.08 percentage points. Finally, as for the duration of each investment and country, a negative cubic trend is observed in most estimates, a negative trend which acts on already negative NPV and IRR indicators –keep in mind that investments which end relatively soon (for example, because they begin to generate negative cash flows) recover a higher residual value of the investment. Analysing the three models together, we can conclude that those countries with higher average and volatility of the price differential represent more favourable environments in which to carry out the arbitrage investment –in model *III*, a unit increase either in the mean price differential or in its standard deviation improves the investment by more than 6,500 €.

Multilevel models tend to pay more attention to the fixed part of the model, using random-effects simply to control for unobservable heterogeneity. However, random effects can themselves be values of interest. Model *III* supports the hypothesis that the dummy coefficients of the converter and battery costs may depend on the country analysed –the likelihood-ratio $\chi^2$ test favours these hypotheses compared to that which assumes random intercepts at a country level. These coefficients are estimated as random slopes, but given the dummy character of the variables (converter and battery costs), they end up conditioning the intercept of the mixed model.

**Table 5.** Multilevel mixed-effects linear regression. NPV and IRR by country.

| Explanatory variables | NPV model | | | IRR model | | |
|---|---|---|---|---|---|---|
| | (I) Random intercept | (II) Random intercept Random slopes for inverter costs | (III) Random intercept Random slopes for inverter and battery costs | (I) Random intercept | (II) Random intercept Random slopes for battery costs | (III) Random intercept Random slopes for inverter and battery costs |
| 1 MW Inverter | -73,032.05*** | -71,086.89*** | -72,234.06*** | -2.87*** | -2.71*** | -2.72*** |
| 2 MW Inverter | -18,951.40*** | -15,389.90*** | -25,062.18*** | -1.44*** | -1.12*** | -1.23*** |
| 3 MW Inverter | -1388.15 | 1926.47 | -3905.61 | -0.61** | -0.36*** | -0.43** |
| 5 MW Inverter | -8801.2 | -11,186.24** | -8,343.2 | 0.32 | 0.18 | 0.21 |
| 6 MW Inverter | -23,490.21*** | -29,378.84*** | -21,325.78** | 0.62** | 0.2 | 0.29 |
| 7 MW Inverter | -39,183.82*** | -45,164.26*** | -37,579.72*** | 0.69** | 0.28** | 0.36** |
| 8 MW Inverter | -57,259.25*** | -66,026.46*** | -55,228.03*** | 0.83*** | 0.24* | 0.37** |
| 9 MW Inverter | -76,247.90*** | -85,137.45*** | -69,043.59*** | 1.17*** | 0.48*** | 0.66*** |
| 10 MW Inverter | -98,132.12*** | -106,998.74*** | -88,392.09*** | 1.20*** | 0.46*** | 0.68*** |
| 10 MWh batt. cost (50,000 €/MWh) | 1,383,010.95*** | 1,329,892.24*** | 1,406,362.41*** | 10.38*** | 7.01*** | 7.87*** |
| 10 MWh batt. cost (100,000 €/MWh) | 898,523.06*** | 862,349.06*** | 903,420.66*** | 4.04*** | 2.02*** | 2.47*** |
| 10 MWh batt. cost (150,000 €/MWh) | 431,335.89*** | 407,337.45*** | 427,522.25*** | 1.42*** | 0.14 | 0.38 |
| TAX (%) | -2,706.96*** | -2,139.66** | -3,221.52*** | -0.10** | -0.06* | -0.07* |
| MeanPrice | -474.8 | -1236.0 | -188.2 | 0.08** | 0.03 | 0.04 |
| SDPrice | 993.2 | 1416.81 | -986.0 | -0.05 | 0.02 | -0.01 |
| MeanGap | 7,242.19*** | 6,830.92*** | 6,696.95*** | 0.04 | 0.04 | 0.03 |
| SDGap | 4,379.51*** | 3,255.91*** | 6,519.64*** | 0.24*** | 0.13*** | 0.17*** |
| DUR | -319,980.94*** | -410,328.71*** | -363,124.97*** | -0.71*** | -6.82*** | -6.05*** |
| DUR$^2$ | 19,817.20*** | 32,676.89*** | 26,806.06*** | | 0.76*** | 0.68*** |
| DUR$^3$ | -392.13*** | -877.17*** | -716.68*** | | -0.03*** | -0.02*** |
| Constant | -176,815.98*** | 7458.05 | -102556.91 | -10.81*** | 3.92 | 1.86 |
| sd(random intercept) | 22,956.5*** | | | 0.92*** | | |
| sd(random slope of battery cost) | | 41,713.3*** | 40,743.1*** | | 2.07*** | 2.05*** |
| sd(random slope of inveter cost) | | | 30,528.7*** | | | 0.52*** |
| sd(1 level residual) | 42,778.5*** | 30,364.4*** | 17,320.5*** | 1.86*** | 0.84*** | 0.69*** |
| Nº of observations | 943 | 943 | 943 | 943 | 943 | 943 |
| Log likelihood | -11423.39 | -11,212.23 | -11012.91 | -1,949.4 | -1,372.3 | -1,331.5 |
| AIC | 22,892.8 | 22,470.5 | 22,073.8 | 3,940.9 | 2,790.6 | 2,711.0 |
| BIC | 23,004.3 | 22,582.0 | 22,190.2 | 4,042.7 | 2,902.1 | 2,827.3 |

**Note**: Significance levels: * p<0.1; ** p<0.05; *** p<0.01.

Figure 8 shows the real values and the fixed and random predictions for the NPV and IRR indicators when the price of the battery is 100,000 €/MWh. Graph (a) shows the NPV values and their predictions for the 4 MW converter (one of the best converters in the NPV mixed model), while graph (b) shows the same information for the IRR indicator in the case of the 10 MW converter (the best option in the IRR mixed model). Thus, graph (a) shows how the fixed prediction (linear prediction for the fixed portion of the model) offers only a partial adjustment with respect to the NPV in some countries, the adjustment being much higher when we add to this fixed prediction the contributions based on predicted random effects (full prediction). Effectively, in countries like Portugal, Switzerland or Spain, there are certain unobservable conditions (apparently not related to the respective discount rates) which determine that the full NPV predictions of these countries (which are very close to their real values) are significantly worse than those predicted simply with the fixed portion of the model with the opposite happening in countries like Latvia, Lithuania or Hungary. For its part, graph (b) shows how this unobserved heterogeneity determines that in countries like Poland, Switzerland or Germany and Luxembourg, the full IRR predictions are worse than those predicted by the fixed portion of the IRR model with the opposite happening in countries like UK, Italy, Lithuania or Norway. The case of Norway stands out; in this country the random effects of the 10 MW converter (price 300,000 €) and 10 MWh battery (price 100,000 €/MWh) jointly considered exceed 4 percentage points in the *III*-IRR model; that is, operating with that investment configuration instead of the base investment –4 MW converter (120,000 €) and 10 MWh battery (200,000 €/MWh)– increases Norwegian IRR index by 4.71 percentage points (0.68 percentage points from the fixed portion of the model and 4.03 from the random portion).

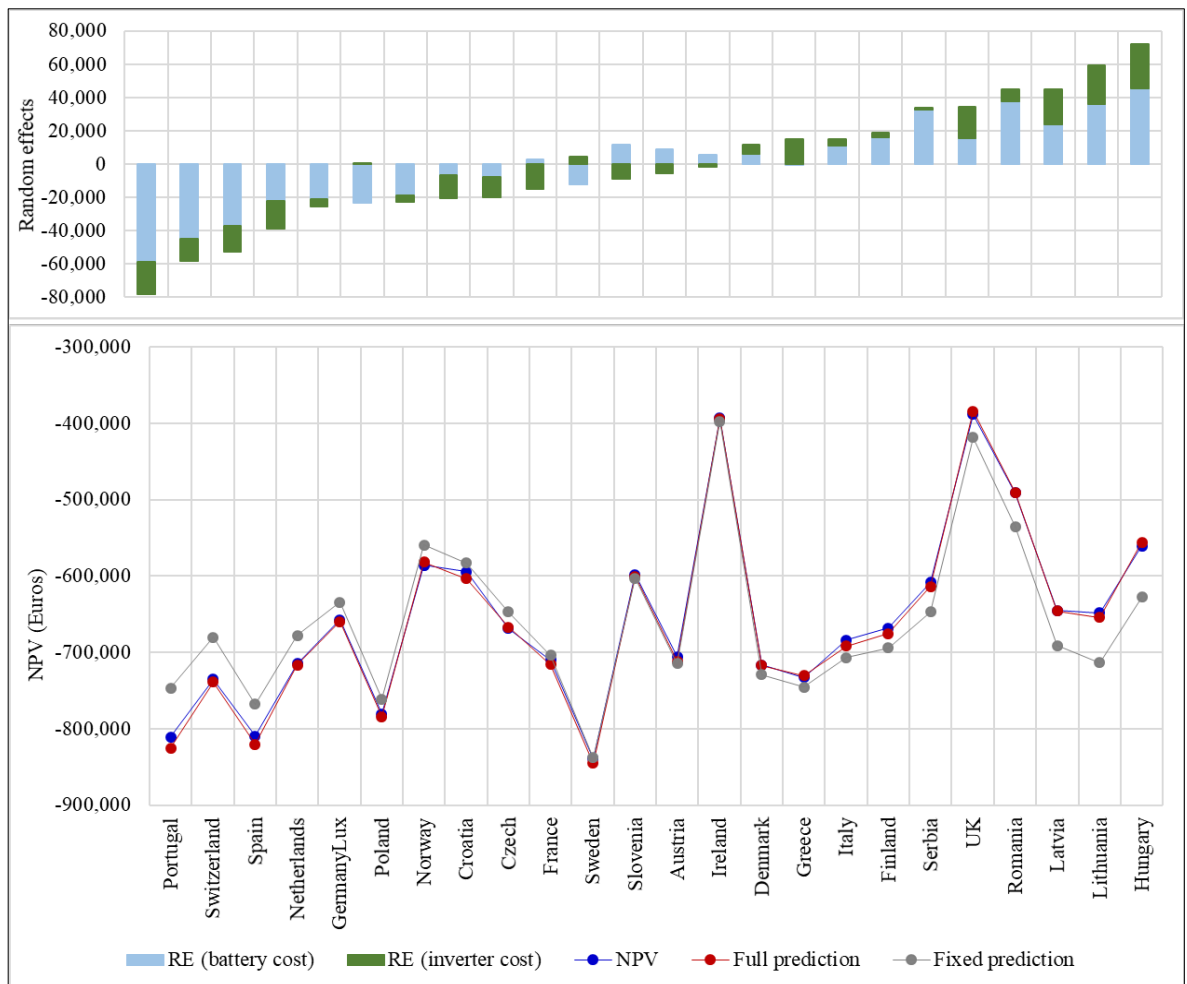

(a) Net Present Value, 4 MW converter, 10 MWh battery ($10^6$ €)

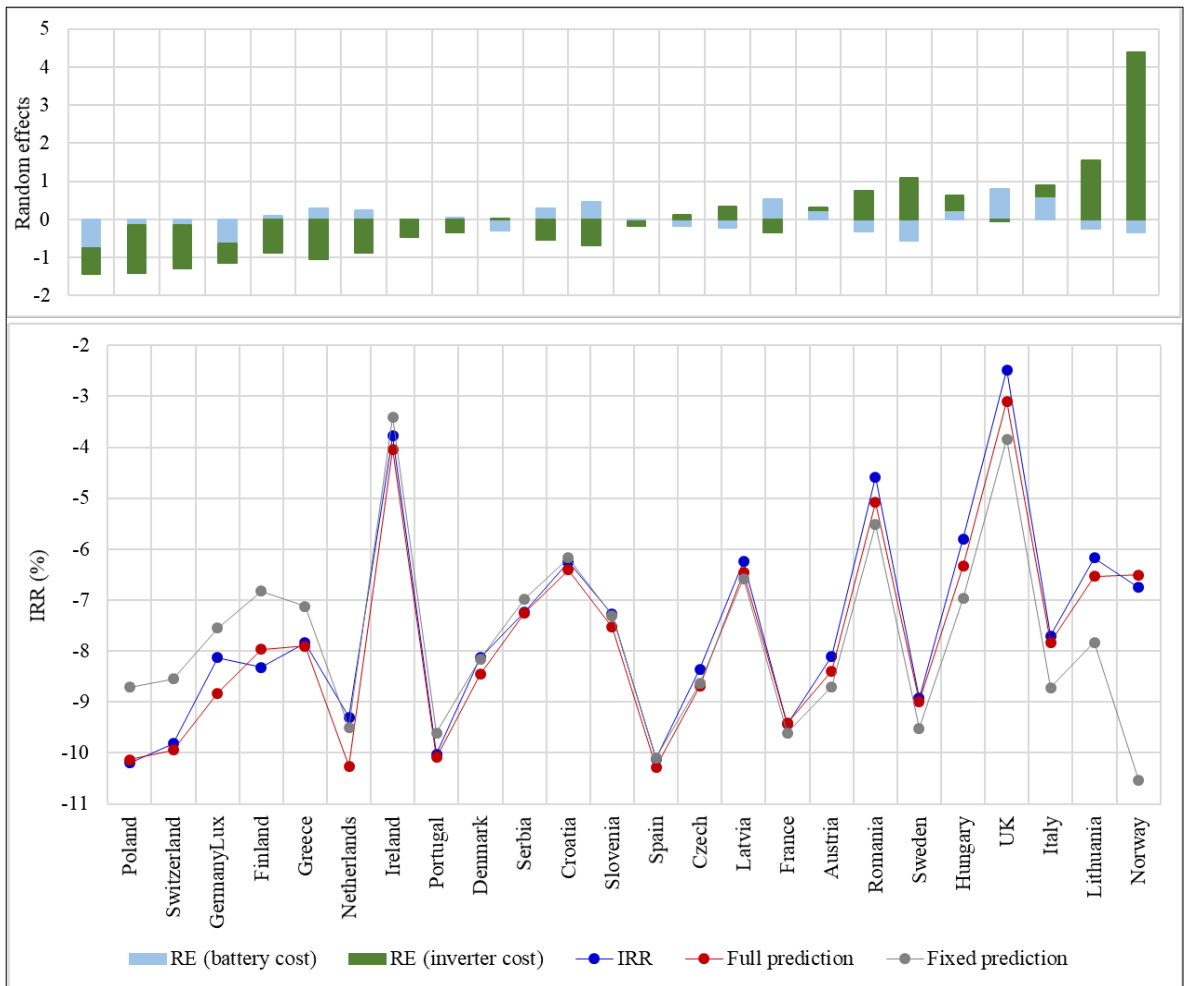

(b) Internal Rate of Return, 10 MW converter, 10 MWh battery ($10^6$ €)

**Figure 8.** Predictions and random effects (RE) by country and investment costs.

The main conclusion of the analysis of the random part of the model is that the existence of unobserved heterogeneity among countries cannot be ignored in these types of international comparisons. In other words, certain stable characteristics of each country (economic, political, fiscal, industrial, etc.) determine that the same investment can generate a significantly different economic return in two different countries, even within the European Union.

**5. Conclusion and business implications.**

In this research, a methodology to compare the business of electricity arbitrage with Li-Ion batteries has been developed and applied to a sample of 24 European countries. To perform a fair business comparison, it is necessary to calculate the tax and risk conditions for each country. To this end, the current methodology used by the Spanish regulator in calculating remuneration rates in the electricity sector has been followed.

Since we are interested in determining the best energy purchase and sale strategy for each country, we developed an optimization framework where, for each country and in a year-by-year sequential way, a mixed-integer linear optimization model is solved by using a Branch-and-Cut algorithm. The model is solved for each national wholesale electricity market with data from the year 2019, and testing different BESS configurations mainly defined by the size of the converter. As a result, for each country, we obtain the purchase and sale decisions that maximize the cash flows until a residual battery capacity level is reached or the business is no longer profitable. The optimal cash-flows resulting from the optimization phase allow us to calculate the financial results, in terms of Net Present Value (NPV) and Internal Rate of Return (IRR), for each possible BESS configuration and country. From there, we focus on the profitability of the business when each country is assigned its most profitable configuration (converter size). Our study concludes by analysing the main (economic and technical) determinants of the profitability of battery arbitrage at the country level, estimating for this a multilevel model which allows the control of unobserved country heterogeneity.

By developing the previously referred methodology, it is possible to answer the research questions that were posed in the introduction of this paper. On the one hand, although arbitrage is not a profitable business for any of the studied countries considering the current Li-Ion battery cost, if battery and converter prices and BESS technology continue the evolution shown in recent years, energy arbitrage will become a profitable business in a few years, for all the countries analysed. On the other hand, there is a great difference in the profitability of the arbitrage business among the analysed countries, with the UK and Ireland being the most attractive markets, and Spain and Portugal the farthest away from profitable results. Finally, the variables which significantly affect the profitability of the business are the cost of the investment (battery plus converter), the tax requirements in each country, and the difference between the maximum and minimum prices of electricity within the day, especially the standard deviation of this price gap. Other variables, such as the discount rate of each country (risk level), or the average level of electricity prices, do not influence the profitability obtained by the business, once we control for unobserved heterogeneity across countries.

Since energy arbitrage in the wholesale market will be a profitable business in the coming years, its scientific analysis is of interest from multiple perspectives (regulatory, academic, business…). The proposed methodology establishes a robust decision support system for this kind of analysis, allowing the study of BESS investments and avoiding improvisations and inefficiencies in decision making.

**References**

Adebayo, A. I., Zamani-Dehkordi, P., Zareipour, H., & Knight, A. M. (2018). Impacts of transmission tariff on price arbitrage operation of energy storage system in Alberta electricity market. *Utilities Policy*, 52, 1-12.

Arcos-Vargas, Á., Canca, D., & Núñez, F. (2020). Impact of battery technological progress on electricity arbitrage: An application to the Iberian market. *Applied Energy*, 260, 114273

Bardo Cáceres, S. (2010). Almacenamiento distribuido en viviendas para alisar la curva de demanda de energía eléctrica. Trabajo final de carrera. Universitat Politècnica de Catalunya (UPC).

Battery University (2019). Section *Learn About Batteries*. Retrieved July, 2020, from https://batteryuniversity.com/learn/

Berrada, A., & Loudiyi, K. (2016). Operation, sizing, and economic evaluation of storage for solar and wind power plants. *Renewable and Sustainable Energy Reviews*, 59, 1117-1129.

Berrada, A., Loudiyi, K., & Zorkani, I. (2016). Valuation of energy storage in energy and regulation markets. *Energy*, 115, 1109-1118.

Bloomberg (2019). *Stock markets capitalisation values*. Retrieved June, 2020, from https://www.bloomberg.com/markets/stocks

Boletín Oficial del Estado (BOE), 2019. Circular 6/2019, de 5 de diciembre, de la Comisión Nacional de los Mercados y la Competencia, por la que se establece la metodología para el cálculo de la retribución de la actividad de distribución de energía eléctrica.

Bradbury, K., Pratson, L., & Patiño-Echeverri, D. (2014). Economic viability of energy storage systems based on price arbitrage potential in real-time U.S. electricity markets. *Applied Energy*, 114, 512-519.

Cameron, A. C., & Trivedi, P. K. (2005). *Microeconometrics: methods and applications*. Cambridge university press.

CleanTechnica (2017). Batteries keep on getting cheaper. Retrieved December 11, 2017, from https://cleantechnica.com/2017/12/11/batteries-keep-getting-cheaper/

Comisión Nacional de los Mercados y la Competencia (CNMC, 2018). *Acuerdo por el que se aprueba la propuesta de metodología de cálculo de la tasa de retribución financiera de la actividad de producción de energía eléctrica a partir de fuentes de energía renovables, cogeneración y residuos para el segundo periodo regulatorio 2020-2025*. Expediente: INF/DE/113/18.

Connolly D., Lund H., Finn P., Mathiesen B.V., & Leahy M. (2011). Practical operation strategies for pumped hydroelectric energy storage (PHES) utilizing electricity price arbitrage. *Energy Policy*, 39, 4189-4196.

Daggett, A., Qadrdan, M., & Jenkins, N. (2017, September). Feasibility of a battery storage system for a renewable energy park operating with price arbitrage. In 2017 IEEE PES Innovative Smart Grid Technologies Conference Europe (ISGT-Europe) (pp. 1-6). IEEE, Torino, Italy.

Das, T., Krishnan, V., & McCalley, J.D. (2015). Assessing the benefits and economics of bulk energy storage technologies in the power grid. *Applied Energy*, 139, 104-118.

Deloitte (2020). *Deloitte International Tax Source*. Retrieved 2020, from https://www.dits.deloitte.com

ENTSO-E's Transparency Platform. Transmission, Day-ahead Prices. Retrieved July, 2020, from https://transparency.entsoe.eu/transmission-domain/r2/dayAheadPrices/show

Dimson, E., Marsh, P. R., Staunton, M., Wilmot, J. J., & McGinnie, P. (2018). *Credit Suisse Global Investment Returns Yearbook 2018*. Credit Suisse Research Institute.

Fernandes, N. (2014). *Finance for Executives: A practical guide for managers*. NPVPublishing.

Goldstein, H. (2011). *Multilevel statistical models*. 4$^{th}$ Ed. John Wiley & Sons.

Gomez-Expósito, A., Sudria, A., Alvarez, E., Díaz, J.L., Pérez-Arriaga, J.I., Arcos-Vargas A., & Pérez de Vargas, J. (2017). El almacenamiento de energía en la distribución eléctrica del futuro. *Real Academia de la Ingeniería*. Madrid.

He, X., Delarue, E., D'haeseleer, W., & Glachant, J. M. (2011). A novel business model for aggregating the values of electricity storage. *Energy Policy*, 39(3), 1575-1585.

Hernández Romero, A. (2016). Análisis económico de un sistema de almacenamiento para la disminución de desvíos de producción en un parque eólico. Trabajo Final de Máster. Escuela Técnica Superior de Ingeniería, Universidad de Sevilla (US).

Hou, P., Enevoldsen, P., Eichman, J., Hu, W., Jacobson, M. Z., & Chen, Z. (2017). Optimizing investments in coupled offshore wind-electrolytic hydrogen storage systems in Denmark. *Journal of Power Sources*, 359, 186-197.

IRENA REthinking Energy (2017). Accelerating the global energy transformation. *International Renewable Energy Agency*, Abu Dhabi.

Jannesar, M. R., Sedighi, A., Savaghebi, M., & Guerrero, J. M. (2018). Optimal placement, sizing, and daily charge/discharge of battery energy storage in low voltage distribution network with high photovoltaic penetration. *Applied Energy*, 226, 957-966.

JOFEMAR Energy (2016). *Almacenamiento electroquímico con baterías de flujo*. Retrieved 2016, from http://www.f2e.es/uploads/doc/20160704075330.f2e_jofemar.pdf.

Kazempour, S. J., Moghaddam, M. P., Haghifam, M. R., & Yousefi, G. R. (2009). Electric energy storage systems in a market-based economy: Comparison of emerging and traditional technologies. *Renewable Energy*, 34 (12), 2630-2639.

Nasrolahpour, E., Kazempour, S. J., Zareipour, H., & Rosehart, W. D. (2016). Strategic sizing of energy storage facilities in electricity markets. *IEEE Transactions on Sustainable Energy*, 7(4), 1462-1472.

OCDE (2020). *Country risk classification*. Retrieved June, 2020, from https://www.oecd.org/trade/topics/export-credits/arrangement-and-sector-understandings/financing-terms-and-conditions/country-risk-classification/

Segui P., C. (2018). *Todo sobre baterías y almacenamiento de energía*. Retrieved June, 2020, from https://www.barriolapinada.es/baterias-almacenamiento-energia/

Sioshansi, R., Denholm, P., Jenkin, T., & Weiss, J. (2009). Estimating the value of electricity storage in PJM: Arbitrage and some welfare effects. *Energy Economics*, 31(2), 269-277.

Terlouw, T., AlSkaifa, T., Bauerb, C. & van Sarka, W. (2019). Multi-objective optimization of energy arbitrage in community energy storage systems using different battery technologies. *Applied Energy*, 239, 356–372.

Trading economics (2020). Retrieved June, 2020, from https://tradingeconomics.com

Vélez Moreno, S. (2012). Estudio de un sistema de almacenamiento de energía eólica por medio de baterías. Trabajo final de carrera. Escuela Técnica Superior de Ingenieros de Minas, Universidad Politécnica de Madrid (UPM).

Walawalkar, R., Apt, J., and Mancini, R. (2007). Economics of electric energy storage for energy arbitrage and regulation in New York. *Energy Policy*, 35, 2558-2568.

Wu, W., & Lin, B. (2018). Application value of energy storage in power grid: A special case of China electricity market. *Energy*, 165, 1191-1199.

Yucekaya, A. (2013). The operational economics of compressed air energy storage systems under uncertainty". *Renewable and Sustainable Energy Reviews*, 22, 298-305.

Zakeri, B., & Syri, S. (2014, May). Economy of electricity storage in the Nordic electricity market: The case for Finland. In 11th International conference on the European energy market (EEM14) (pp. 1-6). IEEE, Krakow, Poland.